\documentclass[aps,twocolumn,showpacs]{revtex4}
\usepackage{graphics,amsmath}
\usepackage{tabularx,graphicx}
\usepackage{epsfig}

\newcommand{\Vel}{V_{\rm el}}
\newcommand{\vF}{v_{\rm F}}
\newcommand{\kF}{k_{\rm F}}
\newcommand{\Vsc}{\delta V_{\rm sc}}
\newcommand{\Ho}{{\cal H}_0}
\newcommand{\Hf}{{\cal H}_{\rm f}}
\newcommand{\EC}{E_{\rm C}}
\newcommand{\DE}{\Delta E}
\begin{document}
\title{Many body effects in finite metallic carbon nanotubes.}
\author{F. Guinea$^{1,2}$}

\affiliation{$^1$Instituto de Ciencia de Materiales de Madrid.
CSIC. Cantoblanco. 28049 Madrid. Spain}
\affiliation{$^2$Department of Physics. Boston University. 590
Commonwealth Avenue, Boston, Massachusetts 02215, USA}


\begin{abstract}
The non homogeneity  of the charge distribution in a carbon
nanotube leads to the formation of an excitonic resonance, in a
similar way to the one observed in X-ray absorption in metals. As
a result, a positive anomaly at low bias appears in the tunnelling
density of states. This effect depends on the screening of the
electron--electron interactions by metallic gates, and it modifies
the coupling of the nanotube to normal and superconducting
electrodes.
\end{abstract}
 \maketitle {\it Introduction.} The
addition of single electrons to finite carbon nanotubes induces
measurable effects, related to the finite spacing between the
energy levels and to the electrostatic energy associated to the
electron
charge\cite{Tetal97,Betal97,Cetal98,Netal99,Betal00b,NCL00,Petal01,Betal02,CN02}
The standard model for Coulomb blockade\cite{AL91,SET} assumes
that the electrostatic potential inside the system is raised by
the addition of an individual electron, preventing tunnelling
unless different charge states are degenerate. The potential is
supposed to be constant throughout the sample. When it is not the
case, non equilibrium effects may occur\cite{UG91,Betal00}, which
are related to the Fermi edge singularities associated to the
sudden ejection of a core electron in a metal\cite{ND69}. One
dimensional systems, like the nanotubes, are good candidates for
observing these effects, as screening is suppressed, and the
electrostatic potential inside them can be modulated by metallic
gates in their proximity. Other effects related to modulations in
the electrostatic potential were considered in\cite{OBH00}.

We calculate in this paper the changes in the tunnelling density
of states due to the non homogeneity of the electrostatic
potential, for different possible experimental setups. The effects
discussed here look similar, but are different from the Luttinger
liquid features expected near a contact\cite{KF92,EG97,KBF97,E99},
which have also been observed in nanotubes\cite{Betal99,Betal01c}.

The next section describes the model to be studied. The main
results are presented next. Then, we analyze how to incorporate
Luttinger liquid effects. The expected behavior at energies
comparable to the spacing between individual electronic levels is
discussed next. The paper concludes with a section on possible
experimental consequences of the effects analyzed here.

{\it The model.} We consider only a nearest neighbor hopping, $t$,
between $\pi$ orbitals at the C atoms. Each subband in a zigzag
nanotube can be modelled by a one dimensional tight binding
hamiltonian with two sites per unit cell and two different
hoppings, $t$ and $t ( 1 + e^{i \phi_n} ) = 2 t e^{i \phi_n / 2}
\cos ( \phi_n / 2 )$, where $\phi_n = ( 2 \pi n ) / N , n = 1 ,
\cdots , N - 1$ is the transverse momentum associated to the
subband, and $N$ is the number of C atoms in a transverse section
of the nanotube.

\begin{figure}
\resizebox{5cm}{!}{\includegraphics[]{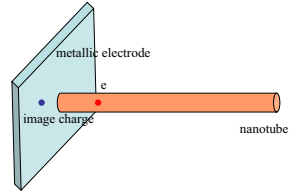}} \caption{Sketch
of one of the geometries considered in order to calculate the
charge distribution in the nanotube.} \label{sketch}
\end{figure}

In a metallic nanotube, $N = 3 M$, where $n$ is an integer. There
are two gapless bands, characterized by $n = \pm M$. Other bands
with $n \ne M$ have a gap, $\Delta_n = t | 1 \pm 2 \cos ( \phi_n /
2 ) |$. Thus, the electronic states of the two bands which cross
the Fermi level can be described by the simple hamiltonian:
\begin{equation}
{\cal H} = t \sum_{n,s} c^\dag_{n,s} c_{n+1,s} + {\rm h. c.}
\label{hamil} \end{equation} The Fermi velocity can be written as
$v_{\rm F} = 2 t a$, where $a$ is the lattice constant. We
consider a finite nanotube with $L$ unit cells. When an electron
hops into the nanotube its charge will be distributed throughout
its length. In a first approximation, this charge can be
calculated using the Hartree or Hartree-Fock approximation. As the
electrostatic potential is  weakly screened in a one dimensional
geometry, the charge distribution can be influenced by metallic
gates in the vicinity. We first consider the setup depicted in
Fig.[\ref{sketch}], where the electron is emitted from a metallic
electrode at one end of the nanotube. More complicated geometries
will be studied later.

We consider the two spin degenerate bands which cross the Fermi
level. The polarization of the remaining bands by the potential
associated to the charge of a single electron will be small, and
it can be treated perturbatively. The radius of the nanotube acts
as a short distance cutoff of the Coulomb potential. We describe
the effects of the metallic electrode by an image charge induced
by the physical charge on the nanotube. We approximate the
electrostatic interaction between electrons located at unit cells
$n_1$ and $n_2$ as measured from the gate, as:
\begin{eqnarray}
V_{\rm tot} ( n_1 , n_2 ) &= &\Vel ( n_1 - n_2 ) - \Vel \left( n_1
+ n_2  + \frac{d}{a} \right) \nonumber \\ &- &\Vel \left( - n_1 -
n_2 +
\frac{d}{a} \right) + \Vel ( - n_1 + n_ 2 ) \nonumber \\
\Vel ( n ) &= &\frac{e^2}{\sqrt{n^2 a^2 + R^2}} \label{potential}
\end{eqnarray}
where $d$ is equal to twice the distance between the nanotube and
the electrode, and $R$ is the radius of the nanotube.

{\it Results.} The electrostatic potential and the induced charge
when the number of electrons in one of the four subbands at the
Fermi level is one above half filling (which is taken as the
neutral situation) is shown in Fig.[\ref{potential}].
\begin{figure}
\resizebox{6cm}{!}{\includegraphics[]{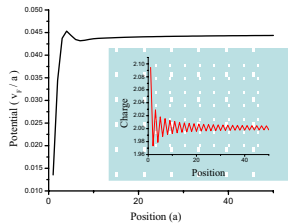}}
\caption{Electrostatic potential and charge (inset) at the edge of
a nanotube with an additional electron and $L = 1024$ unit cells.
The parameters used are $e^2 / \vF = 5.4$, $R/a = 3$ and $d/a =
1$. } \label{potential}
\end{figure}

The calculations have been done using the Hartree approximation.
The results are significantly changed if exchange is included. We
find enhanced Friedel oscillations\cite{SPB00}, and a sizable gap
pinned at the Fermi level. Some of these effects are due to
changes in bulk properties which are unrelated to the addition of
electrons. In the following, we present results obtained within
the Hartree approximation, where the features associated to single
charges are easier to isolate.

A non negligible fraction of the charge is localized by its image
near the electrode. This leads to a reduction of the repulsive
electrostatic potential in that region. This effect is relatively
small compared with the gaps of the higher lying subbands of a
small carbon nanotube, so that the assumption of neglecting their
polarization is justified. We also find substantial Friedel
oscillations, which have periodicity two, as $\kF = \pi / a$.

\begin{figure}
\resizebox{5cm}{!}{\includegraphics[]{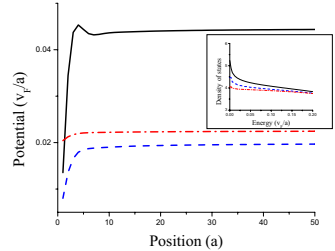}}
\caption{Electrostatic potential induced by a single electron when
there is one electrode (full line), two symmetrically placed
electrodes (dashed line), and no electrodes (dash-dotted line).
The inset shows the non equilibrium density of states for the same
three cases.} \label{electrodes}
\end{figure}

The charge distribution depends strongly on the location of the
external electrodes. Fig.[\ref{electrodes}] compares the potential
calculated previously with the one obtained when there are two
symmetrically placed electrodes at each end of the nanotube, and
in the absence of electrodes, all other parameters being the same.

The effects associated with the non homogeneous charge
distribution after the injection of a single electron are
described, within the Hartree approximation used above, as a
change of all electronic levels. Taking this effect into account,
the electronic density of states is changed as: i) The overlap
between the initial and final eigenstates is reduced. In the limit
of vanishing level spacing, this leads to the orthogonality
catastrophe induced by the sudden switching of a local
potential\cite{A67}. ii) The potential shifts the electrons
towards the electrode, enhancing the density of states at the
Fermi level in its vicinity. This is the so called excitonic
effect\cite{ND69,M91} and it opposes the orthogonality
catastrophe.

The final local density of states, in the limit of vanishing level
splitting, goes as $D ( \omega ) \propto \omega^{[(1-\delta)^2
-1]}$\cite{ND69}, where $\delta > 0$. Hence, for sufficiently
small perturbations, $\delta < 1$, the tunnelling density of
states is enhanced at the ends of the nanotube. The effective
density of states can be obtained by assuming that the hamiltonian
after the charging of the nanotube is modified from an initial
expression $\Ho$ to $\Hf = \Ho + \Vsc$, where $\Vsc$ is the
modification of the Hartree-Fock potential induced by the extra
electron. Assuming that the tunnelling takes place at position 1
and subband 1 near the edge of the nanotube, the tunnelling
density of states at zero temperature is:
\begin{widetext}
\begin{equation}
G ( \omega ) = \prod_{j \ne 1} \left| \,_{f,j}\langle 0 , N | 0 ,
N \rangle_{0,j} \right|^2  
\times  \sum_n \left| \,_{f,1}\langle n , N+1 | c^\dag_0 | 0 , N
\rangle_{0,1} \right|^2 \delta ( \omega - \epsilon_n + \epsilon_0
) 
\label{dos}
\end{equation} \end{widetext}where the states $ | n , N \rangle_{(0,f),j}$ are
eigenstates of ${\cal H}^j_{0,{\rm f}}$ with $N$ electrons, and
$j=1, \cdots 4$ is the band index. As these states are Slater
determinants, the calculation is reduced to a sum of determinants
of overlaps between one particle wavefunctions.

\begin{figure}
\resizebox{6cm}{!}{\includegraphics[]{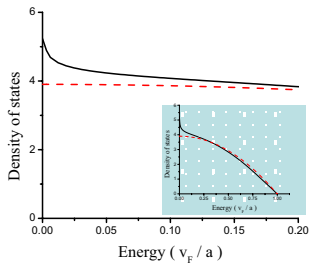}} \caption{Effective
tunneling density of states at the end of a nanowire with $L=1024$
unit cells, using the same parameters as in
Fig.[\protect{\ref{potential}}]. The dotted line is the density of
states in the absence of final state effects. The inset gives the
density of states over the entire bandwidth of the conduction
band.} \label{fig_dos}
\end{figure}

Results obtained for the injection in a neutral nanotube with $L =
1024$ unit cells and the parameters which lead to the
electrostatic potentials shown in Fig.[\ref{electrodes}] are shown
in the inset of Fig.[\ref{electrodes}], and, in more detail, in
Fig.[\ref{fig_dos}], where they are also compared with the results
obtained neglecting $\Vsc$.

{\it Luttinger liquid effects.} The calculation discussed above
takes into account, in an approximate way, the interaction
corrections expected at the boundary of a Luttinger liquid. The
anomalous exponent in the density of states can be estimated using
first order perturbation theory, and it is finite within
Hartree-Fock theory\cite{Metal00}. Note that the calculations also
describe approximately the crossover to a Fermi liquid behavior at
sufficiently high energies\cite{Metal00}.

For sufficiently long wires, we can use the bosonization approach.
The calculation of the final state effects leading to
eq.(\ref{dos}) can be extended without too much difficulty to the
case of a Luttinger liquid with forward interactions. The initial
and final bosonized hamiltonians are\cite{GG04}:
\begin{eqnarray}
\Ho &= &\frac{1}{2} \int_{x>0} d x 4 \pi \vF \Pi^2 ( x ) +
\frac{\vF}{4 \pi} \left[ \partial_x \Phi ( x ) \right]^2 \nonumber
\\ &+ &\frac{e^2}{8 \pi^2} \int_{x>0} dx \int_{y>0} dy \partial_x \Phi ( x )
V_{\rm tot} ( x , y )
\partial_y \Phi ( y ) \nonumber \\
\Hf &= \Ho + &\frac{1}{2 \pi} \int_{x>0} dx \Vsc ( x )
\partial_x \Phi ( x ) \label{hamil_LL} \end{eqnarray}

The hamiltonians $\Ho$ and $\Hf$ differ by a term which is linear
in the boson fields, and there is a simple unitary transformation,
${\cal U}$ which transforms one into the other. The calcuation of
the density of states can be written as the Fourier transform of
the expression:
\begin{eqnarray}
G ( t ) &= &\langle 0 | \psi ( x = 0 ) e^{i {\cal H} t} \psi^\dag
( x = 0 ) | 0 \rangle \nonumber \\ &= &\langle 0 | \psi ( x = 0 )
{\cal U}^{-1} e^{i \Ho t} {\cal U} \psi^\dag (x = 0 ) | 0 \rangle
\label{dos_LL}\end{eqnarray} where $| 0 \rangle $ is the ground
state of $\Ho$, and we are assuming that the end of the nanotube
is at position $x=0$.

The hamiltonians in eq.(\ref{hamil_LL}) are quadratic in the boson
coordinates, and the expression in eq.(\ref{dos_LL}) can be
calculated by normal ordering the operators, which depend
exponentially on the bosonic degrees of freedom. If we neglect the
breakdown of translational symmetry induced by the gates, and the
long range effects of the Coulomb potential, the tunnelling
density of states becomes: \begin{equation} D ( \omega ) \propto |
\omega |^{\frac{1}{g}\left( 1 -g^2 \frac{\delta V}{\vF} \right)^2
- 1} \label{Luttinger}\end{equation} where $g$ is the parameter
which describes the Luttinger liquid properties, and $\delta V = (
2 \pi )^{-1} \lim_{k \rightarrow 0} \int_0^\infty  e^{i k x} \Vsc
( x ) d x$. For repulsive interactions, $g<1$, and Luttinger
liquid effects tend to suppress the positive anomaly studied here.

In the presence of gates, the electron--electron potential is not
translationally invariant. The screening effects of the gates
reduce locally the interactions, which are unscreened away from
the gate. A related situation was considered in\cite{MS95}, a
Luttinger liquid connected to non interacting gates. The quadratic
hamiltonian, eq.(\ref{hamil_LL}), which is defined on a half line,
$x,y > 0$, can be solved by defining a related problem on the
entire axis\cite{LSH01}. We can estimate in a simple way the value
of the parameter $g$ at the edge of the nanotube from the energy
associated to a fluctuation of the charge density on a scale $l<L$
in that region. This estimate, in the thermodynamic limit, leads
to the charge compressibility and to the bulk value of $g$. In the
absence of screening by gates, the inverse compressibility
diverges logarithmically with the size of the system. It is easy
to show that, in the geometry depicted in Fig.[\ref{sketch}], the
energy associated to a fluctuation of size $l$ is given by a term
which diverges logarithmically as $l \rightarrow \infty$ minus a
term which tends to a constant. The second term describes the
screening effects of the gate. Thus, for large nanotubes and at
sufficiently low energies, bulk effects determine the value of
$g$, and the tunnelling density of states will be given by
eq.(\ref{Luttinger}). At high energies, on the other hand, the
screening of the electron--electron interactions by the gate makes
$g \rightarrow 1$ and the non equilibrium effects described here
will dominate. The crossover between these two regimes depends on
the limiting value of the interaction, which, in turn, is
influenced by screening by other parts of the system (see below).

{\it Single electron properties.} At sufficiently low energies or
temperatures, a finite nanotube behaves like a quantum
dot\cite{Tetal97,Betal97,NCL00,Petal01,CN02,Betal02,LBP02,Tetal02}.
For the parameters used here, $e^2 / \vF = 5.4$, the level
splitting and the charging energy are comparable. The Hartree
calculation described above allows us to estimate the charging
energy, $\EC = E_{\rm N+1}+E_{\rm N-1}-2E_{\rm N}$.  The Coulomb
potential considered here, in the absence of gates, leads to $\EC
\approx e^2 L^{-1} \log ( L / R )$. The screening by the vertical
gates, as in Fig.[\ref{sketch}], reduces this value by a quantity
which scales as $L^{-1}$. For the parameters used in
Fig.[\ref{potential}] we find a level spacing (in units of $\vF /
a$) of $\DE = 0.00306$, and a charging energy $\EC = 0.0196$.

The final state effects considered here manifest themselves, at
the scale of the level spacing, as a dependence of the tunnelling
amplitude on the charge state. This effect can be included by
adding assisted hopping terms to the
hamiltonian\cite{G03,SG04,BG04} (see also\cite{H93}). This effect,
in our case, is rather small. Using the parameters in
Fig.[\ref{potential}], the value of the square of the renormalized
creation operator (see eq.(\ref{dos})) at zero energy is 0.002616
for the transition $N \rightarrow N+1$ and 0.002612 for $N+1
\rightarrow N+2$ (note that these transitions involve the same
electronic level). The smallness of the final state effects on
single particle levels shows that the enhancement of the effective
tunnelling of states is a collective phenomenon, which arises from
the cumulative changes in many levels near the Fermi energy.

{\it Conclusions.} We have analyzed, semi--quantitatively, the
appearance of a low energy peak in the tunnelling of states in
nanotubes due to the inhomogeneous potential induced by the charge
of a single electron. This zero bias anomaly is closely related to
the excitonic resonance which appears in X-ray absorption in
metals\cite{ND69,M91}. The formation of this resonance enhances
the tunnelling density of states, although it is induced by the
Coulomb interaction. The strength of this anomaly depends on the
screening of the nanotube by metallic contacts and gates, and,
presumably, it can be tuned experimentally.

The strength of the enhancement of the tunnelling density of
states can be suppressed by Luttinger liquid effects, for
repulsive electron--electron interactions. Luttinger liquid
effects are reduced when the interaction is screened, which, on
the other hand, favors the existence of the excitonic resonance.
The effects discussed here should also be present in multiwall
nanotubes. In a system with many bands at the Fermi level, only
one contributes to the resonance, while the others induce an
orthogonality catastrophe (note that we have considered four bands
here, see eq.(\ref{dos})).

The presence of an excitonic resonance makes the transport
properties of the nanotube similar to those of Luttinger liquids
with attractive interactions. It can lead to the enhancement of
the proximity effect in nanotubes attached to superconducting
electrodes\cite{Ketal99,Metal99,Ketal03}, and increase, in
general, the tendency towards superconductivity in these
systems\cite{G01}.

{\it Acknowledgements.} I am thankful to Boston University, for
its kind hospitality. I have benefited from useful conversations
with A. Castro-Neto, C. Chamon, J. Gonz\'alez, M. Grifoni, and L.
Levitov.
\bibliography{Coulomb_blockade}

\begin{thebibliography}{38}
\expandafter\ifx\csname natexlab\endcsname\relax\def\natexlab#1{#1}\fi
\expandafter\ifx\csname bibnamefont\endcsname\relax
  \def\bibnamefont#1{#1}\fi
\expandafter\ifx\csname bibfnamefont\endcsname\relax
  \def\bibfnamefont#1{#1}\fi
\expandafter\ifx\csname citenamefont\endcsname\relax
  \def\citenamefont#1{#1}\fi
\expandafter\ifx\csname url\endcsname\relax
  \def\url#1{\texttt{#1}}\fi
\expandafter\ifx\csname urlprefix\endcsname\relax\def\urlprefix{URL }\fi
\providecommand{\bibinfo}[2]{#2}
\providecommand{\eprint}[2][]{\url{#2}}

\bibitem[{\citenamefont{Tans et~al.}(1997)\citenamefont{Tans, Devoret, Dai,
  Thess, Smalley, Geerligs, and Dekker}}]{Tetal97}
\bibinfo{author}{\bibfnamefont{S.~J.} \bibnamefont{Tans}},
  \bibinfo{author}{\bibfnamefont{M.~H.} \bibnamefont{Devoret}},
  \bibinfo{author}{\bibfnamefont{H.}~\bibnamefont{Dai}},
  \bibinfo{author}{\bibfnamefont{A.}~\bibnamefont{Thess}},
  \bibinfo{author}{\bibfnamefont{R.~E.} \bibnamefont{Smalley}},
  \bibinfo{author}{\bibfnamefont{L.~J.} \bibnamefont{Geerligs}},
  \bibnamefont{and} \bibinfo{author}{\bibfnamefont{C.}~\bibnamefont{Dekker}},
  \bibinfo{journal}{Nature} \textbf{\bibinfo{volume}{386}},
  \bibinfo{pages}{474} (\bibinfo{year}{1997}).

\bibitem[{\citenamefont{Bockrath et~al.}(1997)\citenamefont{Bockrath, D,
  McEuen, Chopra, Zettl, Thess, and Smalley}}]{Betal97}
\bibinfo{author}{\bibfnamefont{M.}~\bibnamefont{Bockrath}},
  \bibinfo{author}{\bibfnamefont{H.~C.} \bibnamefont{D}},
  \bibinfo{author}{\bibfnamefont{P.~L.} \bibnamefont{McEuen}},
  \bibinfo{author}{\bibfnamefont{N.~G.} \bibnamefont{Chopra}},
  \bibinfo{author}{\bibfnamefont{A.}~\bibnamefont{Zettl}},
  \bibinfo{author}{\bibfnamefont{A.}~\bibnamefont{Thess}}, \bibnamefont{and}
  \bibinfo{author}{\bibfnamefont{R.~E.} \bibnamefont{Smalley}},
  \bibinfo{journal}{Science} \textbf{\bibinfo{volume}{275}},
  \bibinfo{pages}{1922} (\bibinfo{year}{1997}).

\bibitem[{\citenamefont{Cobden et~al.}(1998)\citenamefont{Cobden, Bockrath,
  McEuen, Rinzler, and Smalley}}]{Cetal98}
\bibinfo{author}{\bibfnamefont{D.~H.} \bibnamefont{Cobden}},
  \bibinfo{author}{\bibfnamefont{M.}~\bibnamefont{Bockrath}},
  \bibinfo{author}{\bibfnamefont{P.~L.} \bibnamefont{McEuen}},
  \bibinfo{author}{\bibfnamefont{A.~G.} \bibnamefont{Rinzler}},
  \bibnamefont{and} \bibinfo{author}{\bibfnamefont{R.~E.}
  \bibnamefont{Smalley}}, \bibinfo{journal}{Phys. Rev. Lett.}
  \textbf{\bibinfo{volume}{81}}, \bibinfo{pages}{681} (\bibinfo{year}{1998}).

\bibitem[{\citenamefont{Nyg{\aa}rd et~al.}(1999)\citenamefont{Nyg{\aa}rd,
  Cobden, Bockrath, McEuen, and Lindelof}}]{Netal99}
\bibinfo{author}{\bibfnamefont{J.}~\bibnamefont{Nyg{\aa}rd}},
  \bibinfo{author}{\bibfnamefont{D.~H.} \bibnamefont{Cobden}},
  \bibinfo{author}{\bibfnamefont{M.}~\bibnamefont{Bockrath}},
  \bibinfo{author}{\bibfnamefont{P.}~\bibnamefont{McEuen}}, \bibnamefont{and}
  \bibinfo{author}{\bibfnamefont{P.}~\bibnamefont{Lindelof}},
  \bibinfo{journal}{Appl. Phys.} \textbf{\bibinfo{volume}{69}},
  \bibinfo{pages}{297} (\bibinfo{year}{1999}).

\bibitem[{\citenamefont{Bachtold et~al.}(2000)\citenamefont{Bachtold, Fuhrer,
  Plyasunov, Forero, Anderson, Zettl1, and McEuen}}]{Betal00b}
\bibinfo{author}{\bibfnamefont{A.}~\bibnamefont{Bachtold}},
  \bibinfo{author}{\bibfnamefont{M.~S.} \bibnamefont{Fuhrer}},
  \bibinfo{author}{\bibfnamefont{S.}~\bibnamefont{Plyasunov}},
  \bibinfo{author}{\bibfnamefont{M.}~\bibnamefont{Forero}},
  \bibinfo{author}{\bibfnamefont{E.~H.} \bibnamefont{Anderson}},
  \bibinfo{author}{\bibfnamefont{A.}~\bibnamefont{Zettl1}}, \bibnamefont{and}
  \bibinfo{author}{\bibfnamefont{P.~L.} \bibnamefont{McEuen}},
  \bibinfo{journal}{Phys. Rev. Lett.} \textbf{\bibinfo{volume}{84}},
  \bibinfo{pages}{6082} (\bibinfo{year}{2000}).

\bibitem[{\citenamefont{Nyg{\aa}rd et~al.}(2000)\citenamefont{Nyg{\aa}rd,
  Cobden, and Lindelof}}]{NCL00}
\bibinfo{author}{\bibfnamefont{J.}~\bibnamefont{Nyg{\aa}rd}},
  \bibinfo{author}{\bibfnamefont{D.~H.} \bibnamefont{Cobden}},
  \bibnamefont{and} \bibinfo{author}{\bibfnamefont{P.~E.}
  \bibnamefont{Lindelof}}, \bibinfo{journal}{Nature}
  \textbf{\bibinfo{volume}{408}}, \bibinfo{pages}{342} (\bibinfo{year}{2000}).

\bibitem[{\citenamefont{Postma et~al.}(2001)\citenamefont{Postma, Teepen, Yao,
  Grifoni, and Dekker}}]{Petal01}
\bibinfo{author}{\bibfnamefont{H.~W.~C.} \bibnamefont{Postma}},
  \bibinfo{author}{\bibfnamefont{T.}~\bibnamefont{Teepen}},
  \bibinfo{author}{\bibfnamefont{Z.}~\bibnamefont{Yao}},
  \bibinfo{author}{\bibfnamefont{M.}~\bibnamefont{Grifoni}}, \bibnamefont{and}
  \bibinfo{author}{\bibfnamefont{C.}~\bibnamefont{Dekker}},
  \bibinfo{journal}{Science} \textbf{\bibinfo{volume}{293}},
  \bibinfo{pages}{76} (\bibinfo{year}{2001}).

\bibitem[{\citenamefont{Buitelaar et~al.}(2002)\citenamefont{Buitelaar,
  Bachtold, Nussbaumer, Iqbal, and Sch\"onenberger}}]{Betal02}
\bibinfo{author}{\bibfnamefont{M.~R.} \bibnamefont{Buitelaar}},
  \bibinfo{author}{\bibfnamefont{A.}~\bibnamefont{Bachtold}},
  \bibinfo{author}{\bibfnamefont{T.}~\bibnamefont{Nussbaumer}},
  \bibinfo{author}{\bibfnamefont{M.}~\bibnamefont{Iqbal}}, \bibnamefont{and}
  \bibinfo{author}{\bibfnamefont{C.}~\bibnamefont{Sch\"onenberger}},
  \bibinfo{journal}{Phys. Rev. Lett.} \textbf{\bibinfo{volume}{88}},
  \bibinfo{pages}{156801} (\bibinfo{year}{2002}).

\bibitem[{\citenamefont{Cobden and Nyg{\aa}rd}(2002)}]{CN02}
\bibinfo{author}{\bibfnamefont{D.~H.} \bibnamefont{Cobden}} \bibnamefont{and}
  \bibinfo{author}{\bibfnamefont{J.}~\bibnamefont{Nyg{\aa}rd}},
  \bibinfo{journal}{Phys. Rev. Lett.} \textbf{\bibinfo{volume}{89}},
  \bibinfo{pages}{046803} (\bibinfo{year}{2002}).

\bibitem[{\citenamefont{Averin and Likharev}(1991)}]{AL91}
\bibinfo{author}{\bibfnamefont{D.~V.} \bibnamefont{Averin}} \bibnamefont{and}
  \bibinfo{author}{\bibfnamefont{K.~K.} \bibnamefont{Likharev}}, in
  \emph{\bibinfo{booktitle}{Mesoscopic Phenomena in Solids}}, edited by
  \bibinfo{editor}{\bibfnamefont{B.~L.} \bibnamefont{Altshuler}},
  \bibinfo{editor}{\bibfnamefont{P.~A.} \bibnamefont{Lee}}, \bibnamefont{and}
  \bibinfo{editor}{\bibfnamefont{R.~A.} \bibnamefont{Webb}}
  (\bibinfo{publisher}{Elsevier, Amsterdam}, \bibinfo{year}{1991}).

\bibitem[{\citenamefont{Grabert and Devoret}(1992)}]{SET}
\bibinfo{editor}{\bibfnamefont{H.}~\bibnamefont{Grabert}} \bibnamefont{and}
  \bibinfo{editor}{\bibfnamefont{M.~H.} \bibnamefont{Devoret}}, eds.,
  \emph{\bibinfo{title}{Single Electron Tunneling}}
  (\bibinfo{publisher}{Plenum, New York}, \bibinfo{year}{1992}).

\bibitem[{\citenamefont{Ueda and Guinea}(1991)}]{UG91}
\bibinfo{author}{\bibfnamefont{M.}~\bibnamefont{Ueda}} \bibnamefont{and}
  \bibinfo{author}{\bibfnamefont{F.}~\bibnamefont{Guinea}},
  \bibinfo{journal}{Z. Phys. B: Condens. Matter} \textbf{\bibinfo{volume}{85}},
  \bibinfo{pages}{413} (\bibinfo{year}{1991}).

\bibitem[{\citenamefont{Bascones et~al.}(2000)\citenamefont{Bascones, Herrero,
  Guinea, and Sch\"on}}]{Betal00}
\bibinfo{author}{\bibfnamefont{E.}~\bibnamefont{Bascones}},
  \bibinfo{author}{\bibfnamefont{C.~P.} \bibnamefont{Herrero}},
  \bibinfo{author}{\bibfnamefont{F.}~\bibnamefont{Guinea}}, \bibnamefont{and}
  \bibinfo{author}{\bibfnamefont{G.}~\bibnamefont{Sch\"on}},
  \bibinfo{journal}{Phys. Rev. B} \textbf{\bibinfo{volume}{61}},
  \bibinfo{pages}{16778} (\bibinfo{year}{2000}).

\bibitem[{\citenamefont{Nozi\`eres and di~Dominicis}(1969)}]{ND69}
\bibinfo{author}{\bibfnamefont{P.}~\bibnamefont{Nozi\`eres}} \bibnamefont{and}
  \bibinfo{author}{\bibfnamefont{C.}~\bibnamefont{di~Dominicis}},
  \bibinfo{journal}{Phys. Rev.} \textbf{\bibinfo{volume}{178}},
  \bibinfo{pages}{1097} (\bibinfo{year}{1969}).

\bibitem[{\citenamefont{Oreg et~al.}(2000)\citenamefont{Oreg, Byczuk, and
  Halperin}}]{OBH00}
\bibinfo{author}{\bibfnamefont{Y.}~\bibnamefont{Oreg}},
  \bibinfo{author}{\bibfnamefont{K.}~\bibnamefont{Byczuk}}, \bibnamefont{and}
  \bibinfo{author}{\bibfnamefont{B.~I.} \bibnamefont{Halperin}},
  \bibinfo{journal}{Phys. Rev. Lett.} \textbf{\bibinfo{volume}{85}},
  \bibinfo{pages}{365} (\bibinfo{year}{2000}).

\bibitem[{\citenamefont{Kane and Fisher}(1992)}]{KF92}
\bibinfo{author}{\bibfnamefont{C.~L.} \bibnamefont{Kane}} \bibnamefont{and}
  \bibinfo{author}{\bibfnamefont{M.~P.~A.} \bibnamefont{Fisher}},
  \bibinfo{journal}{Phys. Rev. Lett.} \textbf{\bibinfo{volume}{68}},
  \bibinfo{pages}{1220} (\bibinfo{year}{1992}).

\bibitem[{\citenamefont{Egger and Gogolin}(1997)}]{EG97}
\bibinfo{author}{\bibfnamefont{R.}~\bibnamefont{Egger}} \bibnamefont{and}
  \bibinfo{author}{\bibfnamefont{A.~O.} \bibnamefont{Gogolin}},
  \bibinfo{journal}{Phys. Rev. Lett.} \textbf{\bibinfo{volume}{79}},
  \bibinfo{pages}{5082} (\bibinfo{year}{1997}).

\bibitem[{\citenamefont{Kane et~al.}(1997)\citenamefont{Kane, Balents, and
  Fisher}}]{KBF97}
\bibinfo{author}{\bibfnamefont{C.}~\bibnamefont{Kane}},
  \bibinfo{author}{\bibfnamefont{L.}~\bibnamefont{Balents}}, \bibnamefont{and}
  \bibinfo{author}{\bibfnamefont{M.~P.~A.} \bibnamefont{Fisher}},
  \bibinfo{journal}{Phys. Rev. Lett.} \textbf{\bibinfo{volume}{79}},
  \bibinfo{pages}{5086} (\bibinfo{year}{1997}).

\bibitem[{\citenamefont{Egger}(1999)}]{E99}
\bibinfo{author}{\bibfnamefont{R.}~\bibnamefont{Egger}},
  \bibinfo{journal}{Phys. Rev. Lett.} \textbf{\bibinfo{volume}{83}},
  \bibinfo{pages}{5547} (\bibinfo{year}{1999}).

\bibitem[{\citenamefont{Bockrath et~al.}(1999)\citenamefont{Bockrath, Cobden,
  Lu, Rinzler, Smalley, Balents, and McEuen}}]{Betal99}
\bibinfo{author}{\bibfnamefont{M.}~\bibnamefont{Bockrath}},
  \bibinfo{author}{\bibfnamefont{D.~H.} \bibnamefont{Cobden}},
  \bibinfo{author}{\bibfnamefont{J.}~\bibnamefont{Lu}},
  \bibinfo{author}{\bibfnamefont{A.~G.} \bibnamefont{Rinzler}},
  \bibinfo{author}{\bibfnamefont{R.~E.} \bibnamefont{Smalley}},
  \bibinfo{author}{\bibfnamefont{L.}~\bibnamefont{Balents}}, \bibnamefont{and}
  \bibinfo{author}{\bibfnamefont{P.~L.} \bibnamefont{McEuen}},
  \bibinfo{journal}{Nature} \textbf{\bibinfo{volume}{397}},
  \bibinfo{pages}{598} (\bibinfo{year}{1999}).

\bibitem[{\citenamefont{Bachtold et~al.}(2001)\citenamefont{Bachtold, de~Jonge,
  Grove-Rasmussen, McEuen, Buitelaar, and Schö\"onenberger}}]{Betal01c}
\bibinfo{author}{\bibfnamefont{A.}~\bibnamefont{Bachtold}},
  \bibinfo{author}{\bibfnamefont{M.}~\bibnamefont{de~Jonge}},
  \bibinfo{author}{\bibfnamefont{K.}~\bibnamefont{Grove-Rasmussen}},
  \bibinfo{author}{\bibfnamefont{P.~L.} \bibnamefont{McEuen}},
  \bibinfo{author}{\bibfnamefont{M.}~\bibnamefont{Buitelaar}},
  \bibnamefont{and}
  \bibinfo{author}{\bibfnamefont{C.}~\bibnamefont{Schö\"onenberger}},
  \bibinfo{journal}{Phys. Rev. Lett.} \textbf{\bibinfo{volume}{87}},
  \bibinfo{pages}{166801} (\bibinfo{year}{2001}).

\bibitem[{\citenamefont{Sablikov et~al.}(2000)\citenamefont{Sablikov, Polyakov,
  and B\"uttiker}}]{SPB00}
\bibinfo{author}{\bibfnamefont{V.~A.} \bibnamefont{Sablikov}},
  \bibinfo{author}{\bibfnamefont{S.~V.} \bibnamefont{Polyakov}},
  \bibnamefont{and}
  \bibinfo{author}{\bibfnamefont{M.}~\bibnamefont{B\"uttiker}},
  \bibinfo{journal}{Phys. Rev. B} \textbf{\bibinfo{volume}{61}},
  \bibinfo{pages}{13763} (\bibinfo{year}{2000}).

\bibitem[{\citenamefont{Anderson}(1967)}]{A67}
\bibinfo{author}{\bibfnamefont{P.~W.} \bibnamefont{Anderson}},
  \bibinfo{journal}{Phys. Rev.} \textbf{\bibinfo{volume}{164}},
  \bibinfo{pages}{352} (\bibinfo{year}{1967}).

\bibitem[{\citenamefont{Mahan}(1991)}]{M91}
\bibinfo{author}{\bibfnamefont{G.~D.} \bibnamefont{Mahan}},
  \emph{\bibinfo{title}{Many Particle Physics}} (\bibinfo{publisher}{Plenum,
  New York}, \bibinfo{year}{1991}).

\bibitem[{\citenamefont{Meden et~al.}(2000)\citenamefont{Meden, Metzner,
  Schollw\"ock, Schneider, Stauber, and Sch\"onhammer}}]{Metal00}
\bibinfo{author}{\bibfnamefont{V.}~\bibnamefont{Meden}},
  \bibinfo{author}{\bibfnamefont{W.}~\bibnamefont{Metzner}},
  \bibinfo{author}{\bibfnamefont{U.}~\bibnamefont{Schollw\"ock}},
  \bibinfo{author}{\bibfnamefont{O.}~\bibnamefont{Schneider}},
  \bibinfo{author}{\bibfnamefont{T.}~\bibnamefont{Stauber}}, \bibnamefont{and}
  \bibinfo{author}{\bibfnamefont{K.}~\bibnamefont{Sch\"onhammer}},
  \bibinfo{journal}{Eur. Phys. J. B} \textbf{\bibinfo{volume}{16}},
  \bibinfo{pages}{631} (\bibinfo{year}{2000}).

\bibitem[{\citenamefont{Gonz\'alez and Guinea}(2004)}]{GG04}
\bibinfo{author}{\bibfnamefont{J.}~\bibnamefont{Gonz\'alez}} \bibnamefont{and}
  \bibinfo{author}{\bibfnamefont{F.}~\bibnamefont{Guinea}}
  (\bibinfo{year}{2004}), \eprint{cond-mat/0410387}.

\bibitem[{\citenamefont{Maslov and Stone}(1995)}]{MS95}
\bibinfo{author}{\bibfnamefont{D.~L.} \bibnamefont{Maslov}} \bibnamefont{and}
  \bibinfo{author}{\bibfnamefont{M.}~\bibnamefont{Stone}},
  \bibinfo{journal}{Phys. Rev. B} \textbf{\bibinfo{volume}{52}},
  \bibinfo{pages}{R5539} (\bibinfo{year}{1995}).

\bibitem[{\citenamefont{Levitov et~al.}(2001)\citenamefont{Levitov, Shytov, and
  Halperin}}]{LSH01}
\bibinfo{author}{\bibfnamefont{L.~S.} \bibnamefont{Levitov}},
  \bibinfo{author}{\bibfnamefont{A.~V.} \bibnamefont{Shytov}},
  \bibnamefont{and} \bibinfo{author}{\bibfnamefont{B.~I.}
  \bibnamefont{Halperin}}, \bibinfo{journal}{Phys. Rev. B}
  \textbf{\bibinfo{volume}{64}}, \bibinfo{pages}{075322}
  (\bibinfo{year}{2001}).

\bibitem[{\citenamefont{Liang et~al.}(2002)\citenamefont{Liang, Bockrath, and
  Park}}]{LBP02}
\bibinfo{author}{\bibfnamefont{W.}~\bibnamefont{Liang}},
  \bibinfo{author}{\bibfnamefont{M.}~\bibnamefont{Bockrath}}, \bibnamefont{and}
  \bibinfo{author}{\bibfnamefont{H.}~\bibnamefont{Park}},
  \bibinfo{journal}{Phys. Rev. Lett.} \textbf{\bibinfo{volume}{88}},
  \bibinfo{pages}{126801} (\bibinfo{year}{2002}).

\bibitem[{\citenamefont{Thorwart et~al.}(2002)\citenamefont{Thorwart, Grifoni,
  Cuniberti, Postma, and Dekker}}]{Tetal02}
\bibinfo{author}{\bibfnamefont{M.}~\bibnamefont{Thorwart}},
  \bibinfo{author}{\bibfnamefont{M.}~\bibnamefont{Grifoni}},
  \bibinfo{author}{\bibfnamefont{G.}~\bibnamefont{Cuniberti}},
  \bibinfo{author}{\bibfnamefont{H.~W.~C.} \bibnamefont{Postma}},
  \bibnamefont{and} \bibinfo{author}{\bibfnamefont{C.}~\bibnamefont{Dekker}},
  \bibinfo{journal}{Phys. Rev. Lett.} \textbf{\bibinfo{volume}{89}},
  \bibinfo{pages}{196402} (\bibinfo{year}{2002}).

\bibitem[{\citenamefont{Guinea}(2003)}]{G03}
\bibinfo{author}{\bibfnamefont{F.}~\bibnamefont{Guinea}},
  \bibinfo{journal}{Phys. Rev. B} \textbf{\bibinfo{volume}{67}},
  \bibinfo{pages}{195104} (\bibinfo{year}{2003}).

\bibitem[{\citenamefont{Stauber and Guinea}(2004)}]{SG04}
\bibinfo{author}{\bibfnamefont{T.}~\bibnamefont{Stauber}} \bibnamefont{and}
  \bibinfo{author}{\bibfnamefont{F.}~\bibnamefont{Guinea}},
  \bibinfo{journal}{Phys. Rev. B} \textbf{\bibinfo{volume}{69}},
  \bibinfo{pages}{035301} (\bibinfo{year}{2004}).

\bibitem[{\citenamefont{Borda and Guinea}(2004)}]{BG04}
\bibinfo{author}{\bibfnamefont{L.}~\bibnamefont{Borda}} \bibnamefont{and}
  \bibinfo{author}{\bibfnamefont{F.}~\bibnamefont{Guinea}},
  \bibinfo{journal}{Phys. Rev. B} \textbf{\bibinfo{volume}{70}},
  \bibinfo{pages}{125118} (\bibinfo{year}{2004}).

\bibitem[{\citenamefont{Hirsch}(1993)}]{H93}
\bibinfo{author}{\bibfnamefont{J.}~\bibnamefont{Hirsch}},
  \bibinfo{journal}{Phys. Rev. B} \textbf{\bibinfo{volume}{48}},
  \bibinfo{pages}{3327} (\bibinfo{year}{1993}).

\bibitem[{\citenamefont{Kasumov et~al.}(1999)\citenamefont{Kasumov, Deblock,
  Kociak, Reulet, Bouchiat, Khodos, Gorbatov, Volkov, Journet, and
  Burghard}}]{Ketal99}
\bibinfo{author}{\bibfnamefont{A.~Y.} \bibnamefont{Kasumov}},
  \bibinfo{author}{\bibfnamefont{R.}~\bibnamefont{Deblock}},
  \bibinfo{author}{\bibfnamefont{M.}~\bibnamefont{Kociak}},
  \bibinfo{author}{\bibfnamefont{B.}~\bibnamefont{Reulet}},
  \bibinfo{author}{\bibfnamefont{H.}~\bibnamefont{Bouchiat}},
  \bibinfo{author}{\bibfnamefont{I.~I.} \bibnamefont{Khodos}},
  \bibinfo{author}{\bibfnamefont{Y.~B.} \bibnamefont{Gorbatov}},
  \bibinfo{author}{\bibfnamefont{V.~T.} \bibnamefont{Volkov}},
  \bibinfo{author}{\bibfnamefont{C.}~\bibnamefont{Journet}}, \bibnamefont{and}
  \bibinfo{author}{\bibfnamefont{M.}~\bibnamefont{Burghard}},
  \bibinfo{journal}{Science} \textbf{\bibinfo{volume}{284}},
  \bibinfo{pages}{1508} (\bibinfo{year}{1999}).

\bibitem[{\citenamefont{Morpurgo et~al.}(1999)\citenamefont{Morpurgo, Kong,
  Marcus, and Dai}}]{Metal99}
\bibinfo{author}{\bibfnamefont{A.~F.} \bibnamefont{Morpurgo}},
  \bibinfo{author}{\bibfnamefont{J.}~\bibnamefont{Kong}},
  \bibinfo{author}{\bibfnamefont{C.~M.} \bibnamefont{Marcus}},
  \bibnamefont{and} \bibinfo{author}{\bibfnamefont{H.}~\bibnamefont{Dai}},
  \bibinfo{journal}{Science} \textbf{\bibinfo{volume}{286}},
  \bibinfo{pages}{263} (\bibinfo{year}{1999}).

\bibitem[{\citenamefont{Kasumov et~al.}(2003)\citenamefont{Kasumov, Kociak,
  Ferrier, Deblock, Gu\'eron, Reulet, Khodos, St\'ephan, and
  Bouchiat}}]{Ketal03}
\bibinfo{author}{\bibfnamefont{A.}~\bibnamefont{Kasumov}},
  \bibinfo{author}{\bibfnamefont{M.}~\bibnamefont{Kociak}},
  \bibinfo{author}{\bibfnamefont{M.}~\bibnamefont{Ferrier}},
  \bibinfo{author}{\bibfnamefont{R.}~\bibnamefont{Deblock}},
  \bibinfo{author}{\bibfnamefont{S.}~\bibnamefont{Gu\'eron}},
  \bibinfo{author}{\bibfnamefont{B.}~\bibnamefont{Reulet}},
  \bibinfo{author}{\bibfnamefont{I.}~\bibnamefont{Khodos}},
  \bibinfo{author}{\bibfnamefont{O.}~\bibnamefont{St\'ephan}},
  \bibnamefont{and} \bibinfo{author}{\bibfnamefont{H.}~\bibnamefont{Bouchiat}},
  \bibinfo{journal}{Phys. Rev. B} \textbf{\bibinfo{volume}{68}},
  \bibinfo{pages}{214521} (\bibinfo{year}{2003}).

\bibitem[{\citenamefont{Gonz\'alez}(2001)}]{G01}
\bibinfo{author}{\bibfnamefont{J.}~\bibnamefont{Gonz\'alez}},
  \bibinfo{journal}{Phys. Rev. Lett.} \textbf{\bibinfo{volume}{87}},
  \bibinfo{pages}{136401} (\bibinfo{year}{2001}).

\end{thebibliography}
\end{document}